# Medical Radioisotopes: Supply-chain Resilience, Distributed Production, and Access

**Authors:**

McClain D*, Winklehner D*†, Farhadi F, Alonso JR, Esfahani SA, Huchthausen C, Moon J, Sabet H, Heidari P, Conrad JM

**Affiliations:**

- Laboratory for Nuclear Science, Department of Physics, Massachusetts Institute of Technology, 77 Massachusetts Avenue, Building 26, Cambridge, MA 02139, USA (D McClain, D Winklehner, JR Alonso, JM Conrad, C Huchthausen, J Moon)
- Center for Nuclear Medicine and Theranostic Sciences, Department of Radiology, Massachusetts General Hospital, Boston, MA 02114, USA (SA Esfahani, F Farhadi, H Sabet, P Heidari)
- Lawrence Berkeley National Laboratory, 1 Cyclotron Rd, Berkeley, CA 94720, USA (JR Alonso)

**Correspondence:**

D Winklehner (winklehn@mit.edu)

*These authors contributed equally; †Corresponding author

## Summary

Over 50 million diagnostic procedures involving medical radioisotopes are performed per year and the number of therapeutic and theranostic agents continues to climb. Yet, the global supply chain is far from stable and many regions are underserved. This is the second of a Series of two review papers providing a comprehensive overview of the landscape of medically relevant radioisotopes. This paper focuses on the supply chain. It presents a critical overview of the 2025–2026 status quo, case studies of acute stressors and disruptions they caused, and methods in place to secure a stable supply. A literature-based analysis is paired with expert interviews. We argue that the ageing fleet of nuclear reactors currently used for the centralised production of many isotopes should be complemented by regional accelerator-based centres, alleviating single-point failures and cross-border logistics complications, and increasing access to short-lived isotopes in harder-to-reach regions and low- and middle-income countries.

## Introduction

Nuclear medicine utilises radiopharmaceuticals developed from radioisotopes to perform numerous diagnostic and therapeutic procedures. Due to the short half-lives of many of these radioisotopes, they cannot be easily stockpiled, and robust supply chains must be in place to prevent shortages. These shortages lead to delayed diagnosis and treatment which can result in higher healthcare costs and worse health outcomes.[1] The supply chain for each radioisotope is a unique network through production, regulation, transportation, and patient delivery (Figure 1). Production of radioisotopes begins with the acquisition of stable target materials which are then irradiated at reactors or accelerator facilities. Reactors produce large quantities of neutrons that enable production of neutron-rich radioisotopes. Accelerators, such as cyclotrons and linear accelerators (linacs), predominantly produce neutron-deficient nuclei. Small volumes of medical radioisotopes for preclinical and early clinical trials are pioneered by research institutes and universities. As new pharmaceuticals move to phase three trials, they outgrow these small production laboratories and are picked up by commercial production sites, a handoff that can be stalled by funding and regulatory compliance. Beyond production, radioisotopes must travel quickly to processing facilities and radiopharmacies for labelling and distribution to hospitals and patients. Furthermore, each level of the supply chain relies on a trained workforce facilitated by universities and technical training programs.

Each stage of the supply chain faces its own risk of shortfall which can be exacerbated by international complications.[2] For example, the availability of stable target materials is a limiting factor in the expansion of the field. Historically, much of the enriched stable supply has been sold from United States (US) stockpiles or enriched in Russia. With heightening geopolitical tensions, the US and European Union (EU) have supported expansion in local enrichment.[3–5]

The present review is the second of two papers in a new Series building on the 2024 Lancet Oncology Series on theranostics. The previous Series covered clinical and production perspectives through radiochemistry trends, while also providing a supply chain perspective through the lens of regulatory access, equity in low- and middle-income countries (LMICs), and workforce capacity.[6–9] The first paper of this Series surveys medically important radioisotopes individually, covering their nuclear properties, production methods and clinical status and identifies realistic alternative pathways to improve research and patient access (Moon J and colleagues). This paper expands on the state and stressors on the medical isotope supply chain as well as discussing technologies that could widen access to radioisotopes worldwide.

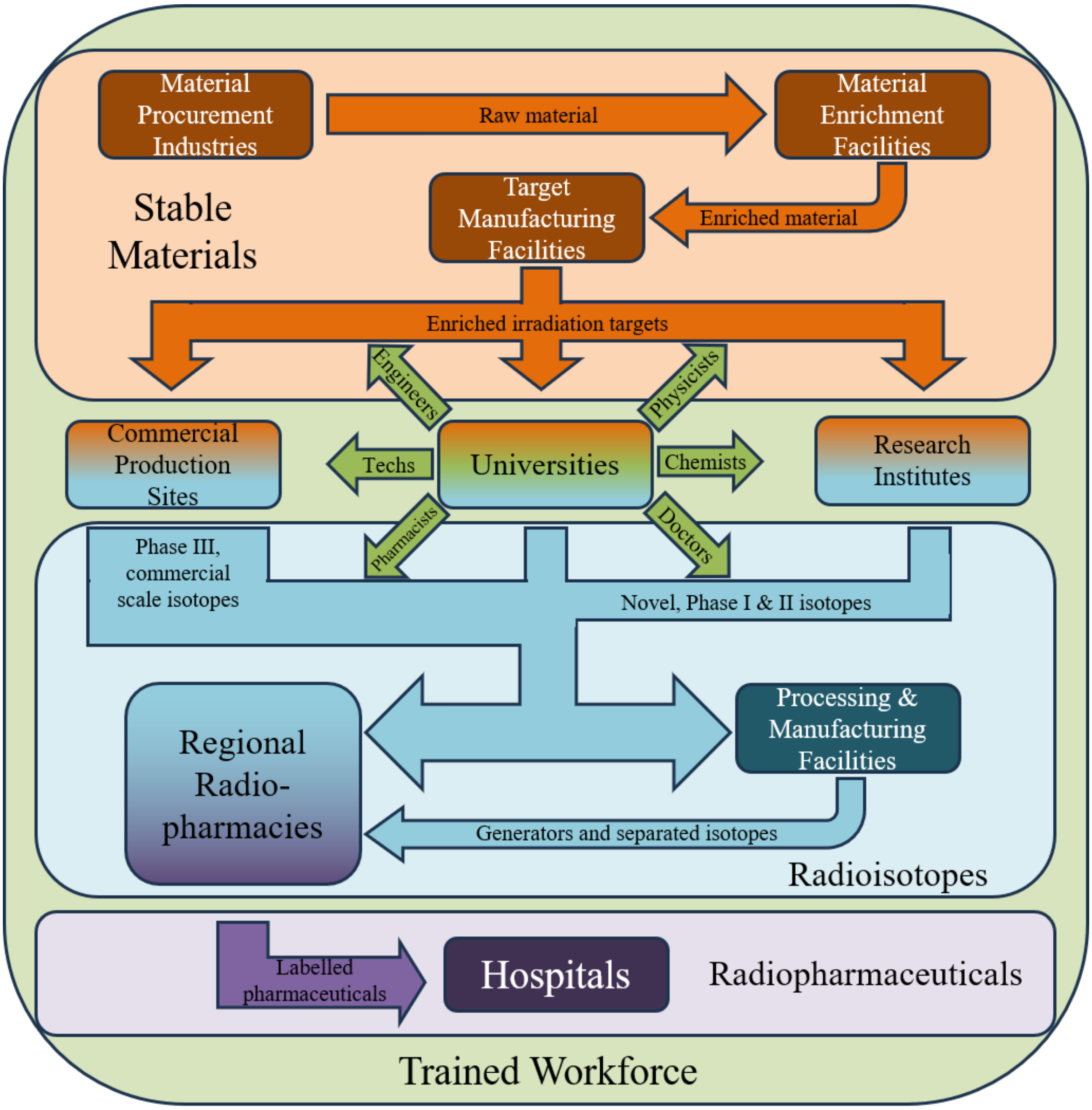

**Figure 1: Simplified modern supply chain network**

### Panel: Search Strategy, Selection Criteria, and Stakeholder Interviews

The review of medical-radioisotope supply chains, distributed production, regulation, and economics drew on both peer-reviewed and grey literature, reflecting that authoritative information on production capacity, policy, and supply-chain stressors is found substantially outside the journal literature. Peer-reviewed sources were identified through Web of Science (all databases), PubMed, and Google Scholar, combining isotope and radiopharmaceutical names with terms including "supply chain," "production capacity," "shortage," "Mo-99," "Lu-177," "Ac-225," "research reactor," "cyclotron," "accelerator," "regulation," "FDA," "NRC," and "EMA." Grey-literature sources included reports from the OECD Nuclear Energy Agency, the IAEA, the US Department of Energy Office of Isotope R&D and Production and its National Isotope Development Center, the US Nuclear Regulatory Commission, the European Medicines Agency, and equivalent national bodies, together with company announcements and regulatory filings documenting recent capacity developments. Recent and current data were prioritised, with emphasis on 2020 onward to capture acute supply-chain stressors (COVID-19 and the war in Ukraine) and the 2025–2026 capacity landscape; older sources were retained where they remain the most authoritative reference.

The case-study and regional-production analyses are additionally informed by semi-structured interviews of 13 stakeholders across national laboratories and academic production centres (5), clinicians (4), and regulatory and oversight bodies (4). Interviewees were identified from North America and Europe as past or present directors or research leads, as well as practicing nuclear medicine clinicians and pharmacists at university hospitals. The interviews covered production, supply, and logistics experienced at different levels with focuses on 2020–2026, the COVID-19 pandemic, and the Russia-Ukraine war. These are presented in aggregate and are shown in italics.

We incorporate non-English sources where production, national regulatory frameworks and supply-chain policy are documented only in the original language and have no English equivalent of record; here, we draw on Russian-, Chinese-, German-, and Japanese-language sources.

## Supply-Chain Fragility and Mitigation Efforts

### The current landscape

While many radioisotopes are made with cyclotrons (in hospitals, universities, and commercial producers such as ARRONAX and Theragenics), linacs (eg Brookhaven National Laboratory, Los Alamos National Laboratory, Niowave), and Rhodotrons (NorthStar Medical Radioisotopes), the bulk of modern-practice radiopharmaceuticals originate from reactors. Reactors around the world produce more than 80% of medically relevant radioisotopes (Figure 2).[5] Much of this volume reflects use of the SPECT workhorse $^{99m}$Tc, the decay product of $^{99}$Mo. $^{99m}$Tc is used in more than 40 million procedures, representing more than 80% of nuclear medicine procedures.[5]

Simultaneous reactor outages in 2010 created a global shortage of $^{99}$Mo. In response, the Organisation for Economic Co-operation and Development Nuclear Energy Agency (OECD-NEA) established the High-Level Group on the Security of Supply of Medical Radioisotopes (HLG-MR) to oversee the supply chain and make recommendations for the future. These recommendations include the establishment of an operating reserve capacity (ORC) at all levels of the supply chain to minimise outage impact, the implementation of full-cost recovery (FCR) to promote production expansion, the phasing-out of highly enriched uranium (HEU) as fuel and targets for non-proliferation, and international consistency through policy and information sharing.[10]

| Project Name | Country | Primary Radioisotopes | Commissioning Year |
|---|---|---|---|
| LVR-15 | Czechia | $^{99}$Mo, $^{153}$Sm, $^{177}$Lu, $^{192}$Ir | 1957 |
| HFR | Netherlands | $^{90}$Y, $^{99}$Mo, $^{161}$Tb, $^{177}$Lu, $^{192}$Ir | 1961 |
| BR-2 | Belgium | $^{99}$Mo, $^{131}$I, $^{177}$Lu, $^{90}$Y | 1961 |
| SAFARI-1 | South Africa | $^{99}$Mo, $^{131}$I, $^{177}$Lu | 1965 |
| MARIA | Poland | $^{32}$P, $^{60}$Co, $^{99}$Mo, $^{131}$I, $^{153}$Sm, $^{166}$Ho, $^{177}$Lu, | 1974 |
| OPAL | Australia | $^{99}$Mo, $^{125}$I, $^{131}$I, $^{177}$Lu | 2006 |

a. Major isotope producing reactors*

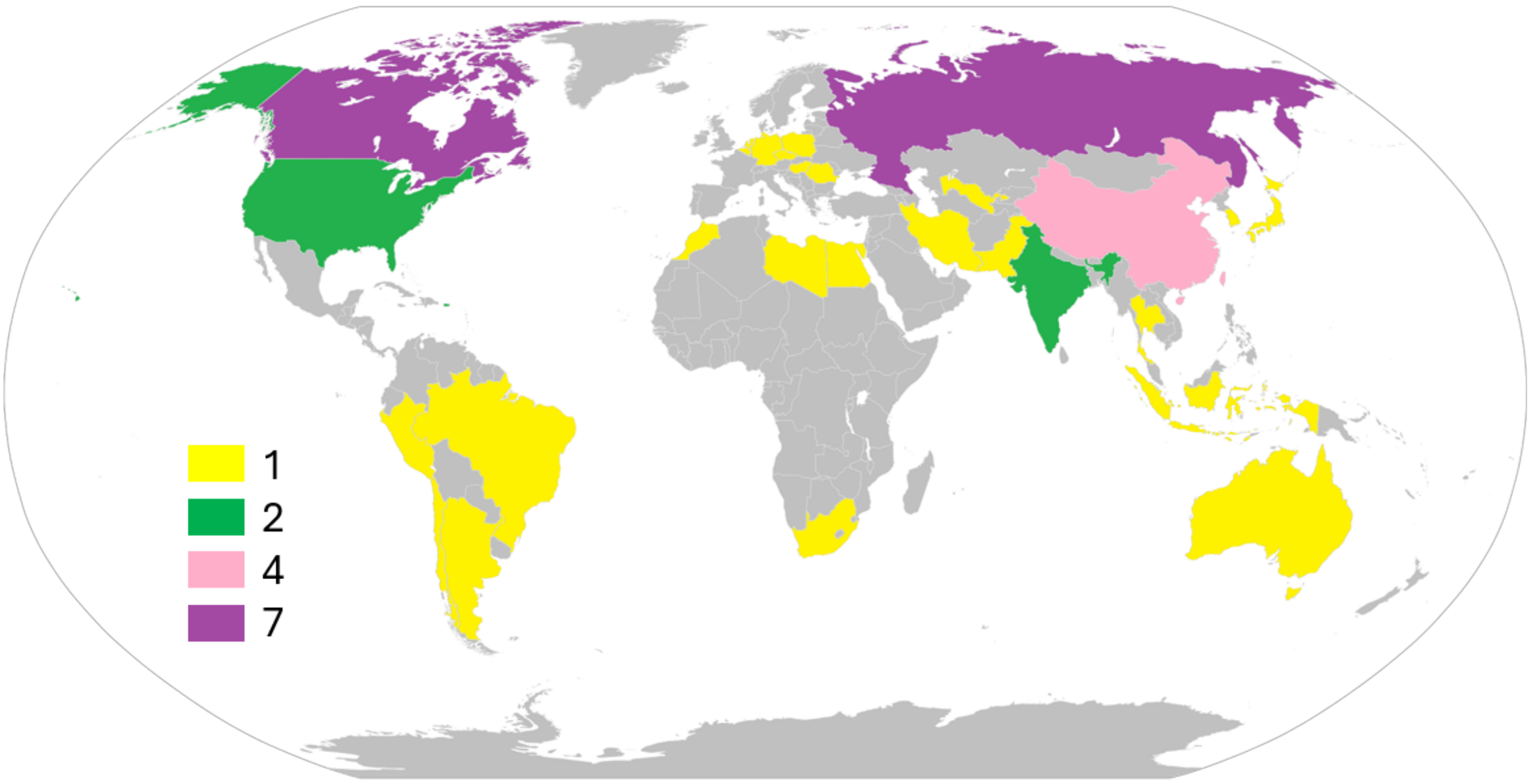

b. Isotope producing reactors by country

*These reactors produce approximately 80% of global nuclear medicine volume

**Figure 2: Medical radioisotope producing reactors**

Data Sources: Research reactors were identified in ref [11], non-research reactors were identified individually, Russian reactors in ref [12], Canadian reactors in ref [13], and Chinese reactors in ref [14]. Estimated production volume comes from ref [5].

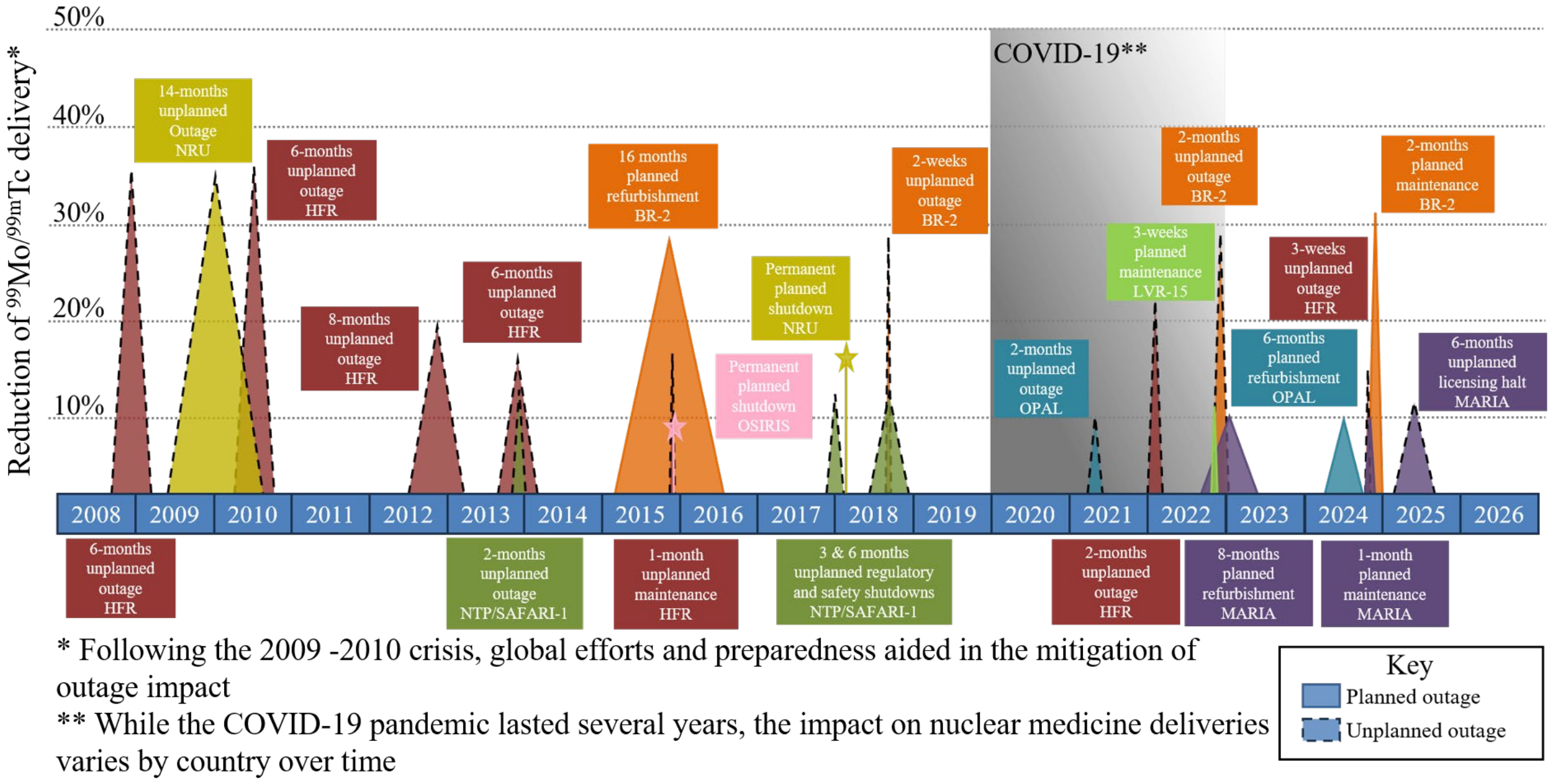

**Figure 3: Timeline of medical radioisotope delivery disruptions**

Data Sources: Impact is derived from ref [5,15–20], event data provided in appendix Table 1.

An ORC of 35% is recommended so production and manufacturing facilities can increase output to mitigate the impact of outages in parallel supply chains. As the major production reactors age, there are accelerating shortages between unplanned stoppages, decommissionings, and necessary refurbishments (Figure 3).

*Even with an established ORC, these reactor outages and the risk associated with reliance on the limited production capacity of ageing reactors remain a forefront concern for interviewed clinicians. During shortages, clinicians, radiopharmacies, and governing bodies often coordinate locally to ration supply and minimise the impact of shortages; meanwhile, hospitals reprioritise exams to focus on the most urgent cases. As such, the impact of short-term events varies by region and ability to mitigate outages.*

Growth in the sector is encouraged by FCR through the inclusion of the full cost of production in the medication price – increasing market competitiveness. The effect of global adherence to FCR remains debated: while promoting production growth, it risks making radioisotopes unaffordable for downstream industry and patients.[5,21] These increased costs would likewise hinder nuclear medicine growth in LMIC.

Beyond production volume, the sector is expanding the variety of available radioisotopes. Research institutions handle small-scale production for radioisotopes in the emerging frontier until the requested volume requires commercial scale production for phase three clinical trials and broad adoption. Risks of gaps in this hand-off are a common concern that can be mitigated by early funding by private producers into their production pathways to be prepared to supplant the small-scale production.

*Interviewees showed concern in these hand-off gaps and stated that $^{225}Ac$, a promising agent for targeted alpha therapy, is currently undergoing these growing pains: the limited number of production sites struggles to support ongoing clinical trials, risking supply chain collapse. Nonetheless, interviewed experts noted that the supply is closely tracked, and production expansion is promising.*

The rollout of these new radiopharmaceuticals can be delayed by a lack of regulatory coordination. Early clinical trials inform governments about radiopharmaceuticals' efficacy and safety, but regulatory acceptance is required before widespread adoption. The lack of international coordination in these approval processes delays global establishment.[8,22] Scaling to commercial production takes significant funding and time and can be delayed further by the regulatory process, as seen with the Pluvicto rollout in 2022–2023.[23]

As new radiopharmaceuticals develop, demand for a skilled workforce increases, outpacing the training of this workforce.[24,25] Coordination among funding agencies, industry, and educational institutes is required to expand workforce training, while expansion of standardised training supports workforce development in LMIC.[6,7,9]

*Workforce was a recurrent issue among interviewees. All levels of the workforce were mentioned with a repeated call for increased training and funding for the production-based workforce – from technicians to PhD researchers – to support the growing industry and supply needs.*

Beyond the chronic stressors of ageing infrastructure, insufficient workforce, and regulatory hurdles, acute stressors, as described below apply further pressure on the supply chain.

## COVID-19

In March 2020, the World Health Organization (WHO) declared COVID-19 a global pandemic. Demand for radiopharmaceuticals decreased as clinical efforts shifted toward COVID-19. Between limited scheduling for nuclear medicine departments, staff reassigned to other departments to combat COVID-19, and patients reluctant to visit clinics, by April 2020, nuclear medicine procedures declined by approximately 50%.[26,27] A separate survey showed a decline of 73% in June 2020 compared to pre-pandemic averages.[28]

Medical radioisotope shipments reliant on international passenger flights were stalled by air travel bans initiated during the pandemic. As essential products, medical radioisotope production and shipment continued; however, significant supply shortage in $^{131}I$ and $^{99m}Tc/^{99}Mo$ generators was reported.[27]

*Contrary to literature, interviewed producers met requested demands and interviewed clinicians did not report widespread supply shortage in US or EU hospitals. The reduction in nuclear medicine procedures made any supply shortage less noticeable.*

Delivery shortages had a greater impact in LMIC without domestic production capacity.[27,29] Due to an inability to import radioisotopes, and community quarantines limiting domestic transportation, some nations' nuclear medicine procedures decreased by as much as 90% in 2020.[26] Nuclear medicine in Asia recovered as early as 2021, before later variants caused further waves of infection.[26] While European nuclear medicine did not experience drastic supply shortages, nuclear medicine procedures did not return to pre-pandemic levels until 2023.[30,31]

## Russia–Ukraine war

In early 2022, following the Russian invasion of Ukraine, nations around the globe placed sanctions on Russia, obstructing supply of target materials for radioisotope production. Proposed legislation would place US sanctions on Russian-produced and enriched isotopes, increasing pressure to replace Russian suppliers.[32]

Russia has been a major supplier of medical radioisotope production material including low-enriched uranium (LEU) and enriched $^{176}$Yb, the precursor for $^{177}$Lu, sparking fears of supply chain collapse for these materials.

*While interviewees stated there were initially some shortages, it was deemed a risk to supply that did not have a large direct impact on radiopharmaceutical availability but nonetheless stimulated western stable isotope production.*

In addition, high-assay low enriched uranium (HALEU) has primarily been produced by downblending HEU from US and Russian stockpiles. As the global initiative of non-proliferation encourages the use of LEU, HALEU is the upper limit of enrichment and thus efficiency in reactors and targets. In 2024, the US signed the Prohibiting Russian Uranium Imports Act into law, banning the import of Russian unirradiated LEU.[33] Nevertheless, Russia supplied 26% of enrichment services to US reactors in 2025.[34]

## Current supply chain resilience efforts

Initiatives and investments to increase domestic production of enriched stable isotopes, enriched uranium, and medically relevant radioisotopes are underway around the world.[3,35] Market pressures have prompted reinvestment in stable isotope enrichment at the US Stable Isotope Production and Research Center (SIPRC) and the Radioisotope Science Center.[5] Likewise, the European Commission introduced a roadmap toward ending Russian energy imports in response to the Russia-Ukraine war, complementing the European Radioisotope Valley Initiative (ERVI) in decreasing foreign dependency of the European nuclear supply chain.[4,36] Further downstream, new large-capacity production facilities are expected in the coming decade, though many such as PALLAS,[15,37] SHINE Chrysalis,[5,16] and the Jules Horowitz Reactor frequently face timeline slippage extending the reliance on ageing infrastructure (appendix Table 3).[5,15]

Beyond production expansions, the supply chain is supported by coordinated supply networks. The European Production of high purity Isotopes by mass Separation for Medical Application (PRISMAP) and the USA National Isotope Development Center (NIDC) operate to create a distribution network for the EU and US respectively, providing an onramp for early research of new radiopharmaceuticals.[38] They connect production sites, research institutes, and hospitals to ensure stable supply of radioisotopes through early clinical trials. By ensuring research-level supply, these networks fast-track innovation toward future life-saving procedures. For established radioisotopes, commercial-scale production organisations navigate distribution independently.

Although national and international distribution arrangements are reported to be adequate, regulatory constraints on international transport could be eased by establishing regional production clusters. Each cluster would serve an area with harmonised regulations and delivery times compatible with the usable lifetimes of the radioisotopes produced.[38] Production clusters are already effective for short-lived PET radioisotopes produced in cyclotrons. With a handful of strategically placed cyclotrons able to deliver their PET radioisotopes within 40–100 km, nations can support their nuclear medicine clinics through regional initiatives.[39]

*European interviewees were generally satisfied with the established cyclotron network in their regions for short-lived radioisotopes. North American respondents were split between satisfaction with production and recommendations for expansion in local production hubs to facilitate distribution to smaller communities.*

# Distributed Accelerator-Based Production

## Production pathways and accelerator landscape

Radioisotopes produced in reactors contain more neutrons than the stable isotopes of these elements and are referred to as "neutron-rich". Due to the high flux of neutrons generated at reactors, these radioisotopes can be made in large quantities, and supply SPECT and therapeutic radioisotopes to large populations with a single production site. PET radioisotopes, on the other hand, are all neutron-deficient, because positron emission ($\beta^+$ decay, where a proton in the radioisotope converts into a neutron) gives the requisite collinear $\gamma$-rays. These radioisotopes are typically produced in cyclotrons, some small enough to be installed in hospitals, eg the automated patient-ready production of $^{18}F$ (fluorodeoxyglucose) for PET imaging. An overview of particle accelerators in use for medical isotope production is given in Figure 4A, and important parameters are listed in Table 1 and appendix Table 4. While linacs and other accelerator types can be employed in specific cases, the work-horse machines for producing neutron-deficient isotopes are cyclotrons delivering up to 1·6 mA beams of 15 to 33 MeV protons. These produce a plethora of valuable radioisotopes, for diagnostic and therapeutic applications.[40] Most of these machines will not easily fit in a hospital but could be installed in distributed regional centres, within convenient shipping distances from nuclear medicine departments. Cyclotrons up to 70 MeV are in regional service (eg ARRONAX). These too can be deployed in regional centres, equipped with hot cells to process the larger volumes of produced radioisotopes into clinical doses.

Accelerator production is smaller, cheaper to maintain, and safer to operate,[41] so installing several production centres can substantially reduce supply-chain problems and mitigate international shipping problems.

While such centres can produce neutron-deficient radioisotopes, today no commercial accelerator can generate a sufficient neutron flux to generate conventionally reactor-produced radioisotopes, such as $^{99}Mo$ and $^{177}Lu$, in clinical quantities. However, as shown in appendix Table 4, photonuclear reactions from electron accelerators are used in the commercial production of $^{225}Ac$ and $^{67}Cu$. Neutron sources under development include deuterium-tritium (D-T) generators driving a subcritical assembly,[42] deuteron linacs, and high-current cyclotrons. These are currently still two to three orders of magnitude below research reactor outputs but require a smaller footprint (Figure 4B, Table 1).

A new family of high-current $H_2^+/D^+$ cyclotrons (HCHC-XX/HCDC-XX) aiming at delivering 10 mA protons or 5 mA deuterons is being developed for beam energies from 1·5 MeV/amu to 60 MeV/amu (100 MeV/amu in reach).[43,44] Simulations suggest that a 3 MeV deuteron version of this cyclotron can produce a thermalisable neutron flux adequate, in principle, for production of a clinically viable $^{99}Mo/^{99m}Tc$ generator through fission of HALEU.[45] Higher energy deuteron cyclotrons could be prolific sources of fast neutrons (10 – 20 MeV or higher) through deuteron breakup. This represents a pathway for medical radioisotope production which could open, for one, efficient production of $^{225}Ac$.[43] This route was demonstrated already for $^{99}Mo$, $^{64}Cu$, and $^{67}Cu$.[46]

Today's supply-chain problems could be mitigated by an adequate number of regional accelerator-based production centres for distribution of radioisotopes for research and clinical use.

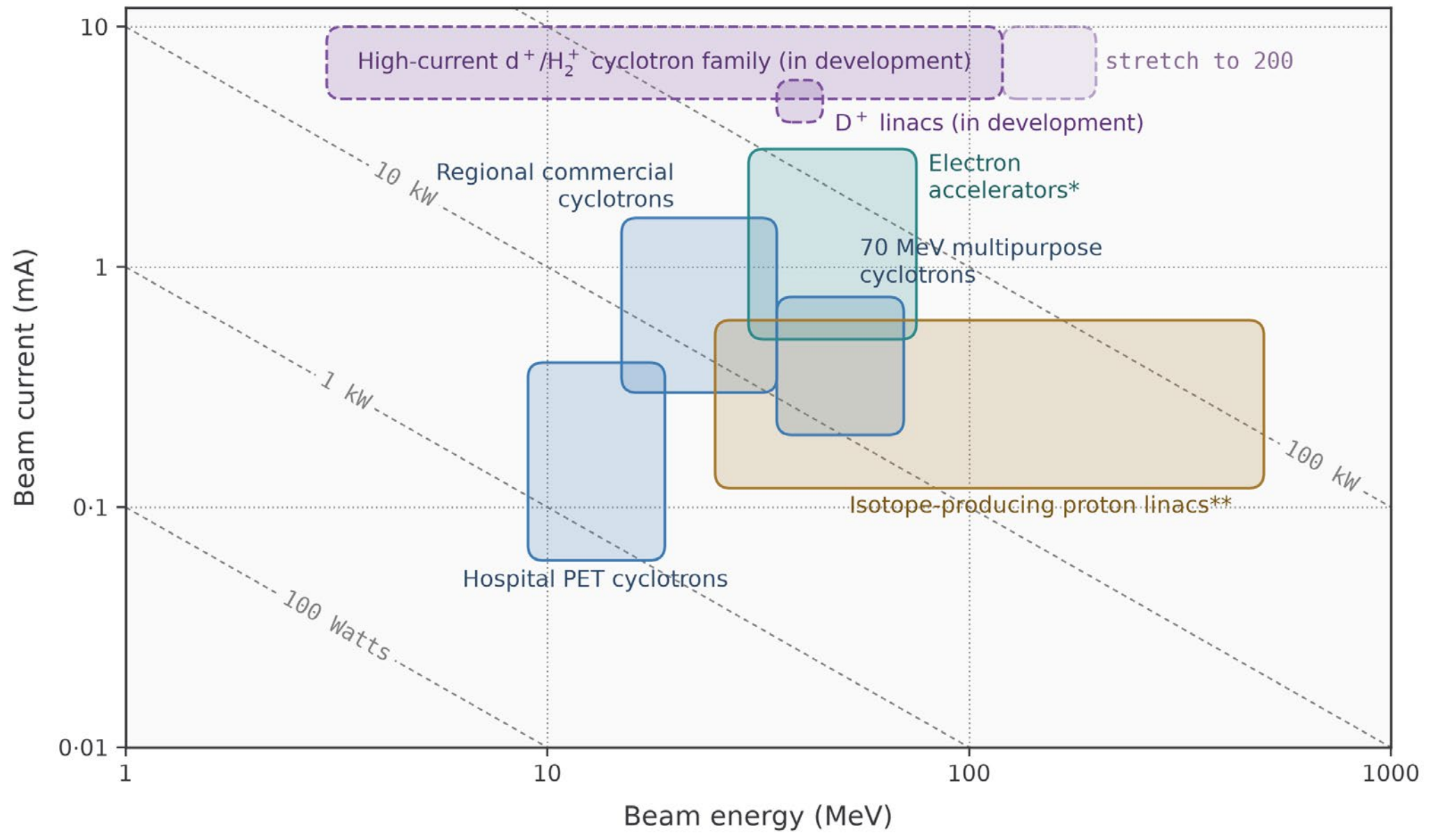

A

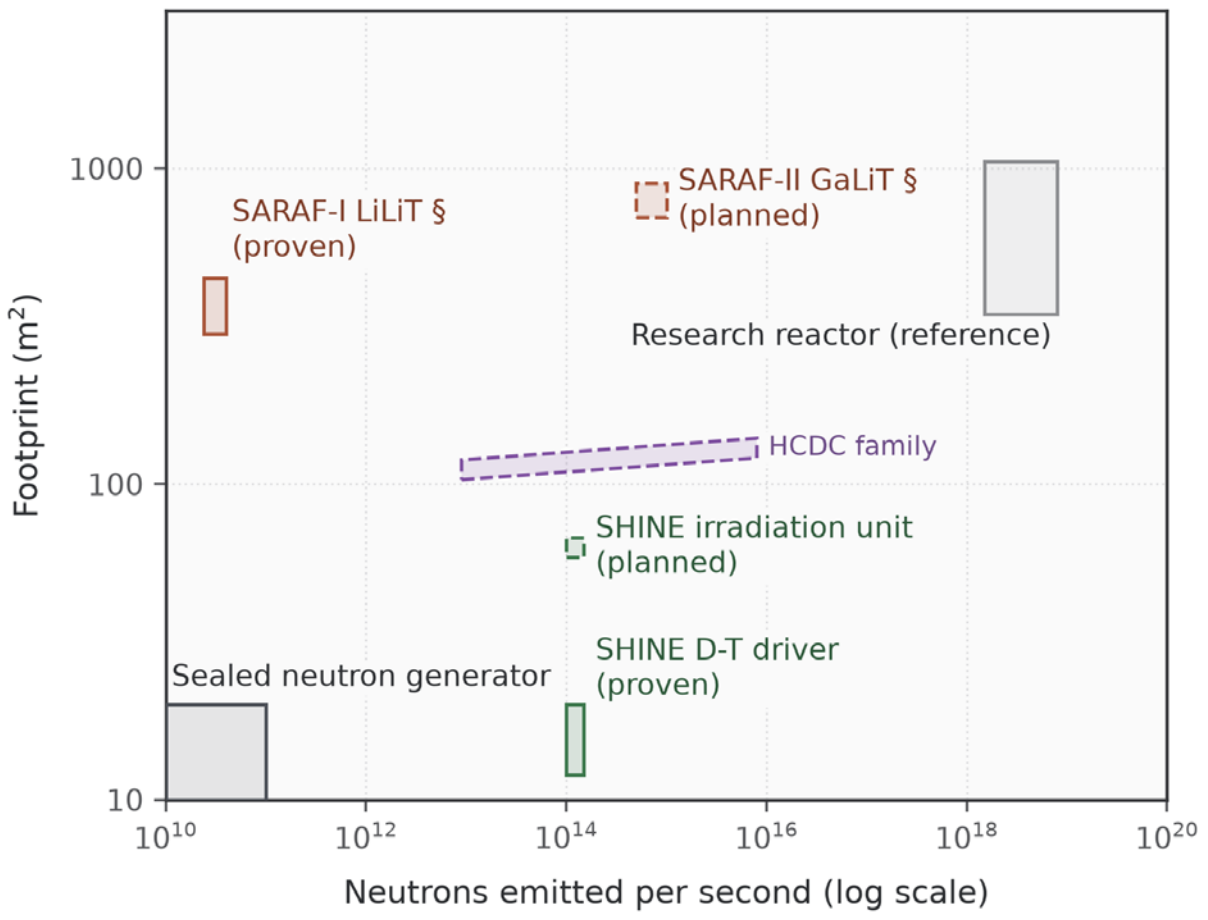

B

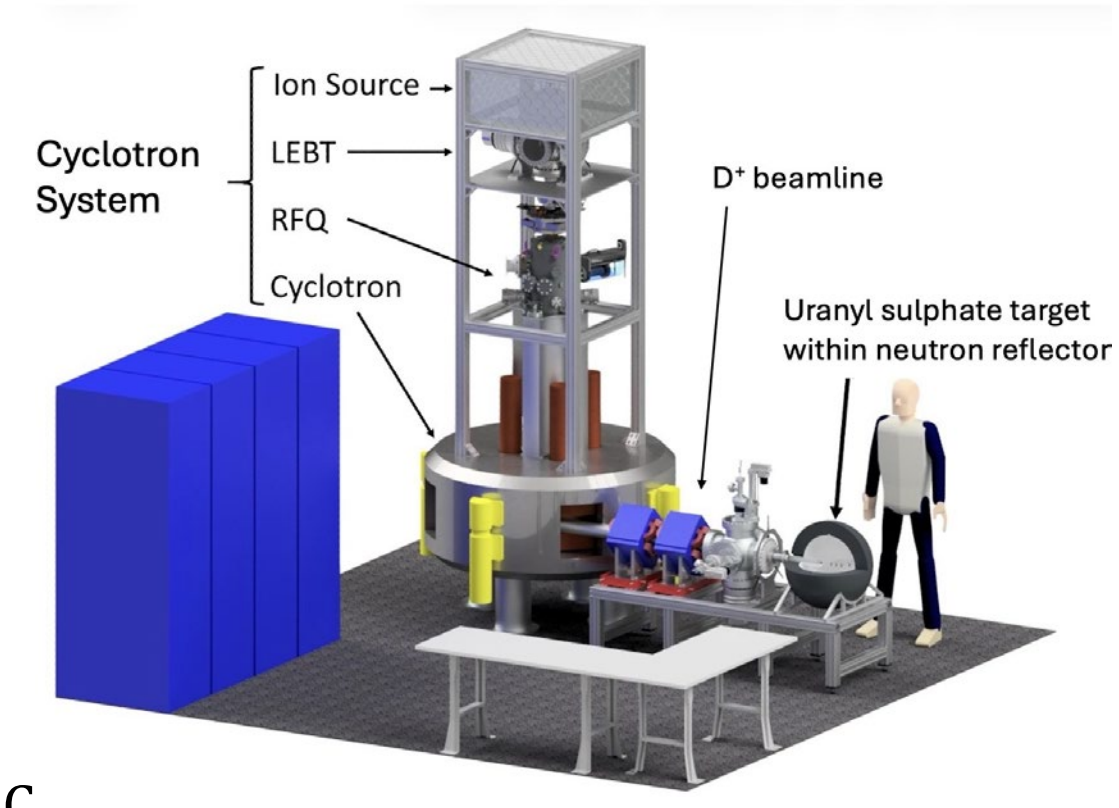

C

**Figure 4: Overview of particle accelerators for medical isotope production. (A) Landscape of beam energy, beam current and beam power. (B) Landscape of neutron production and accelerator footprint. (C) CAD rendering of a 3 MeV cyclotron that can be used for $^{99}$Mo production.[45]**

***Electron accelerators encompass linacs and Rhodotrons.**

****Proton linacs comprise dedicated medical isotope facilities and parasitic usage parallel to basic physics research.**

**§Not currently pursuing medical isotope production but listed as a future application.**

| Machine class | Beam | Footprint‡ | Isotopes Potentially Producible | Production channel examples |
|---|---|---|---|---|
| Low Energy Cyclotrons | 9 – 19 MeV $p^+$ | ~46 $m^2$ | 37 | Direct proton activation, proton-rich isotopes |
| Medium Energy Cyclotrons | 15 – 35 MeV $p^+$ | ~94 $m^2$ | 39 | Direct activation, broader proton-rich portfolio |
| High Energy Cyclotrons | 35 – 70 MeV $p^+$ | ~111 $m^2$ | 40 | Direct activation, spallation-adjacent channels |
| Proton linac (dedicated to isotopes) | 25 – 100 MeV $p^+$ | ~600 $m^2$ | 40 | High-energy activation, $^{225}$Ac from Th |
| Electron accelerators | 30 – 75 MeV $e^-$ | ~72 – 175 $m^2$ | 30 | Photonuclear (γ,n); subcritical fission designs |
| High-current $H_2^+$/$D^+$ cyclotrons (HCHC-XX/HCDC-XX) (concept) | 3 – 120 MeV $d^+$<br>1·5 – 60 MeV $p^+$ | ~110 – 130 $m^2$ | 41 | $d^+$/$p^+$ direct, eg (d,2n), (p,2n)<br>$d^+$ breakup, thermal or fast n |
| $D^+$ linac (SARAF) § | 40 MeV $d^+$ | ~700 $m^2$ | 38 | $d^+$/$p^+$ direct, eg (d,2n), (p,2n)<br>$d^+$ breakup, then (n,p) |
| D-T (SHINE Chrysalis) | 0·3 MeV $d^+$ | ~60 $m^2$ | 20 | $^{235}$U(n,f) |
| Research reactor (reference) | — | ~350–1050 $m^2$ | 27 | Thermal-neutron fission |

**Table 1: Radionuclide production capabilities by device / beam class. Counts are conservative estimates derived from the production pathways discussed in the first paper of this Series (Moon J and colleagues). Isotopes are included if the given device / beam is well matched to a nuclear production pathway capable of yielding quantities and quality suitable for clinical research at minimum. Existing need or commercial competitiveness is not considered here. Direct, secondary-particle, and parent-decay routes are included. Inclusion requires measured or quantitatively supported yields under suitable irradiation and processing conditions. The data sources for energy and footprint are in appendix C.**
**‡Estimates based on the references in appendix C. The footprint includes the accelerator and peripherals (power supplies, shielding, cryoplants, etc.), but not the target stations, radiochemistry, and radiopharmaceutical treatment.**
**§Not currently pursuing medical isotope production but listed as a future application.**

## Clinical Demand and Market Perspectives

Clinical demand for medical radioisotopes now extends well beyond the established workhorses of $^{99m}$Tc, $^{18}$F, $^{131}$I, and $^{68}$Ga that underpin diagnostic imaging and routine therapy. The NETTER-1 and VISION phase 3 trials demonstrated survival benefits with $^{177}$Lu-DOTATATE and $^{177}$Lu-PSMA-617, leading to FDA approvals for Lutathera and Pluvicto and establishing radiopharmaceutical therapy as standard-of-care in somatostatin receptor-positive gastroenteropancreatic neuroendocrine tumours and PSMA-positive metastatic castration-resistant prostate cancer.[47,48] Commercial uptake has paralleled clinical validation: Pluvicto reached US$1·99 billion in FY2025 sales during its third full market year. Lutathera contributed US$816 million. Both agents sustained double-digit growth and moved toward earlier-line disease settings.[49]

Capital markets have responded accordingly. KPMG identified more than 90 radiopharmaceutical transactions between 2021 and 2024, spanning acquisitions, licensing agreements, and equity financings, with aggregate deal value over the preceding decade exceeding US$33 billion.[50] Major transactions include Bristol Myers Squibb's US$4·1 billion acquisition of RayzeBio, Eli Lilly's US$1·4 billion acquisition of POINT Biopharma,[51] and AstraZeneca's acquisition of Fusion Pharmaceuticals for up to US$2·4 billion.[52] This investment pattern reflects a broader reassessment of radiopharmaceuticals as a scalable oncology modality rather than a narrow nuclear medicine niche.

The therapeutic pipeline has expanded in parallel. A systematic analysis of ClinicalTrials.gov from 2014 to 2024 identified 254 therapeutic radiopharmaceutical trials, with annual trial postings increasing by 13·0% per year.[53] $^{177}Lu$-based agents accounted for 52·8% of trials, industry sponsorship reached 49·6%, and prostate cancer together with gastroenteropancreatic neuroendocrine tumours represented 61·8% of disease-specific studies.[53] These data suggest that radiopharmaceutical therapy is no longer confined to isolated academic programmes, but is becoming an organised therapeutic development space with substantial industry participation.

However, this clinical and commercial expansion is approaching a supply-side ceiling. The next generation of therapeutic radioisotopes includes alpha emitters like $^{225}Ac$ and $^{211}At$, matched theranostic pairs like $^{64}Cu/^{67}Cu$ and $^{152}Tb/^{161}Tb$, and emerging beta-emitting platforms like $^{212}Pb$. Because many of these radioisotopes depend on immature, geographically concentrated, low-yield, or single-source production pathways, the limiting step for clinical translation may shift from ligand design or target validation to the availability of radioisotopes at sufficient activity, purity, reproducibility, and regulatory quality for multicentre trials and routine clinical deployment.

## Clinical access and LMIC implications

The argument for distributed production is an argument about access, and nuclear medicine access is highly inhomogeneous. Approximately 27 000 SPECT or SPECT/CT systems are installed across more than 140 countries, against just over 5 200 PET/CT systems across 109 countries; one SPECT camera serves roughly 57 000 people in high-income countries but an average of 33 million people in low-income countries; the corresponding figures for PET are approximately 250 000 and 250 million.[23]

Therapeutic access is scarcer still and narrows sharply for radioisotopes on the emerging frontier. A 105-country IAEA survey reported through the Lancet Oncology Commission on radiotherapy and theranostics found $^{131}I$ therapy for differentiated thyroid cancer is performed in 80 countries at roughly 250 000 procedures annually, $^{177}Lu$-DOTATATE in 39 countries at approximately 43 000 treatments, $^{177}Lu$-PSMA in 32 countries at close to 48 000 treatments, and $^{225}Ac$-PSMA in only 12 countries at approximately 1 000 treatments per year.[23,54] Access contracts by an order of magnitude at each step away from the generator- and reactor-supplied agents that have been in routine practice for decades.

Accelerator-based regional production addresses this geographical access gap directly, but the existing installed base is poorly matched to the need. The IAEA Medical imAGIng and Nuclear mEdicine global resources database (IMAGINE) records cyclotrons in 89 countries, overwhelmingly dedicated to $^{18}F$, and of more than 1 500 cyclotrons installed worldwide by 2023 approximately 1 400 operate below 25 MeV, with only about 130 in the 25–70 MeV range.[23] The deficit is therefore one of energy, current and downstream processing capability rather than machine count. Regional centres with 25–70 MeV accelerators extend the accessible portfolio to $^{68}Ga$, $^{123}I$, $^{64/67}Cu$, $^{89}Zr$ and $^{211}At$, and the high-current deuteron machines discussed above open credible routes to $^{99}Mo$ and $^{225}Ac$.

Where production has been localised, access follows. In India, the Bhabha Atomic Research Centre (BARC) and Board of Radiation and Isotope Technology (BRIT) supply an indigenous $^{177}$Lu-PSMA formulation at approximately 80 000 INR per 3·7 GBq dose – less than one-third the cost of imported product (approximately 250 000 INR) – and distribute ready-to-use $^{177}$Lu radiopharmaceuticals to more than 50 nuclear-medicine centres nationwide.[55,56]

South Africa's NTP Radioisotopes, operating the SAFARI-1 research reactor, achieved the first large-scale, all-LEU production of $^{99}$Mo, eliminating reliance on weapons-grade HEU while securing commercial supply. The facility also became Africa's only source of no-carrier-added $^{177}$Lu, expanding regional access to beta-emitting radioligand therapy.[20,57]

## Regulatory Framework and Commercialisation

When deployed for medical isotope production, accelerators become embedded in the radiopharmaceutical supply chain, subject to nuclear regulatory oversight and pharmaceutical manufacturing regulation. The critical question is not whether an accelerator can produce a radioisotope, but whether the entire system can repeatedly yield a clinically usable radioisotope with validated documentation, regulatory compliance, and traceability from target irradiation to patient administration.

### FDA and EMA approval pathways

In most jurisdictions, radioisotopes produced for human use must satisfy two independent regulatory systems: radiation safety and drug quality. In the US, the Energy Policy Act of 2005 brought accelerator-produced radioactive material under Nuclear Regulatory Commission (NRC) authority which means a high-current cyclotron facility must hold a specific NRC license listing the exact radionuclides it is authorised to produce;[58] adding a new isotope requires a license amendment, a months-long process independent of drug-development timelines. In parallel, the Food and Drug Administration (FDA) regulates radiopharmaceutical manufacturing. A producer must satisfy two agencies with different regulatory considerations. A sterility failure is a current Good Manufacturing Practice (cGMP) problem for the FDA; an airborne contamination event is a radiation safety problem for the NRC; and the two agencies do not coordinate enforcement.

For therapeutic radionuclides such as $^{67}$Cu or $^{225}$Ac, the regulatory burden is heavier than for diagnostic PET tracers. Congress directed the FDA to create streamlined cGMP rules (21 CFR Part 212)[59] for short-lived diagnostic PET drugs, permitting single-person oversight and conditional release before sterility testing is complete. But that streamlined framework is limited to diagnostic agents. Non-PET radiopharmaceuticals fall under standard pharmaceutical cGMP (21 CFR Parts 210 and 211),[60,61] requiring completed sterility verification, second-person checks, and more extensive facility and documentation controls before any batch can be released. The distinction is consequential for operations: PET tracers can be released on parametric controls, while other radiopharmaceuticals must remain under quarantine until sterility is confirmed, compressing the already narrow window between production and administration. The distinction is also infrastructural: a facility built for Part 212 compliance typically cannot be retrofitted for Part 211 without changes to shielding, air handling, and workflow separation.

The EU regulatory architecture is different. Under EudraLex Annex 3, irradiation is considered outside GMP; the radioisotope is classified as starting material, and GMP obligations apply only to subsequent chemical processing, purification, and formulation.[62] A radionuclide producer does not hold a GMP certificate for the irradiation step; the downstream radiopharmaceutical manufacturer must qualify the supplier through audits, specifications, and batch documentation. A US-produced radioisotope qualified under FDA cGMP therefore cannot automatically be transferred into an EU manufacturing chain; its specifications must be accepted by the European manufacturer as equivalent to a starting material produced under a different regulatory philosophy.[63]

### Business models for local accelerator-based facilities

Three organisational models dominate medical isotope production infrastructure, each with distinct strategic advantages. Academic and non-profit regional hubs – exemplified by the NIDC and PRISMAP production networks – are optimal for translational radiochemistry, first-in-human and investigator-initiated trials, workforce training, and production of public-interest radioisotopes lacking near-term commercial viability. A classic economic analysis of academic cyclotron operations demonstrated that such facilities remain financially viable only by integrating research, clinical, and commercial revenue streams, with external radioisotope sales cross-subsidising research capabilities that grant funding alone cannot sustain.[64] For-profit contract development and manufacturing organisations (CDMOs) provide the complementary function: GMP-compliant scale-up, standardised release testing, multi-site trial supply, and commercial distribution. Market incentives, however, tend to concentrate on radioisotopes with established profit margins rather than emerging or low-volume radioisotopes.

A third, increasingly policy-relevant model is the public–private resilience partnership, in which governments co-fund regional production capacity in exchange for contractual reserve-capacity and emergency-supply obligations. Precedents include the European Commission's 2024 approval of €2 billion in Dutch state aid for the PALLAS reactor,[65] justified as securing the supply of essential medicines, and the US Department of Energy's expanded isotope program budget, which couples funding for university and national-laboratory production with conditional loan support for commercial isotope facilities.[35,42]

## Discussion

Deferred procedures caused by supply shortages delay diagnosis and treatment, risking patient outcomes and increasing healthcare costs. Despite mitigation efforts, the supply remains at risk as ageing reactors account for approximately 80% of global nuclear medicine volumes. This reliance is prolonged by slipping timelines of new production (appendix Table 3) and exacerbated by acute global stressors, alongside vulnerabilities within the trained workforce. These stressors, such as the COVID-19 pandemic and the Russia-Ukraine war, pose risks that can only be mitigated via supply diversification and local to global coordination. According to those interviewed for this manuscript, while these geopolitical and pandemic stressors were concerning, they ultimately had minimal impact on North American and European nuclear medicine supply. Meanwhile, reactor outages and workforce constraints remain chronic stressors threatening nuclear medicine.

Particle accelerators have been used for medical isotope production for over half a century, cyclotrons dominating the market. However, they are mainly used for PET and SPECT isotopes, while reactors produce neutron-rich radioisotopes due to their significantly higher neutron flux. Several efforts are underway to bring accelerators closer to the reactor regime. Among them are superconducting deuteron linacs, fusion- or accelerator-driven subcritical assemblies, photonuclear production based on electron machines, and pushes to bring cyclotrons to higher currents (appendix Table 5). No neutron-based efforts currently produce radioisotopes on a commercial scale, and none should be viewed as competitors to nuclear reactors in terms of neutron flux. However, with eg a 5 mA continuous wave deuteron beam, production of neutron-rich isotopes with cyclotrons may soon become feasible, making distributed accelerator centres viable for complementing the existing reactor fleet. The benefits of using many accelerator-based production sites include smaller individual footprints, safer operation, and stability through redundancy, all improving access and resilience. In addition, several therapeutic and theranostic isotopes of increasing interest could be produced with accelerators (eg $^{225}Ac$ and $^{177}Lu$) and a large fleet of such machines would incentivise development of new production pathways. With increased scarcity of emerging radioisotopes in LMICs--with limited domestic production and research capacities-- regional production infrastructure could lower costs and improve isotope availability, as seen for example in South Africa and India.

Differences in regulatory standards can restrict cross-border collaboration by requiring companies to conduct their own inspections to verify quality. Furthermore, approval timelines are determined by the slowest step in the chain – radionuclide license amendment, process validation, facility inspection, and cross-border starting-material qualification – with each regulator reviewing a different segment. For future production centres, regulatory strategy should be planned before facility design. The radioisotope portfolio (diagnostic PET versus therapeutic) determines the applicable cGMP pathway, the pathway determines the facility, and thus the economics.

The public networks, private industry, and public–private partnerships should be understood as complementary rather than competing: reactors remain indispensable for neutron-rich therapeutic isotopes such as $^{177}Lu$, while generators and conventional cyclotrons serve established diagnostic volumes. Additional accelerator-based production hubs can support the established networks by providing regional redundancy and expansion of available radioisotopes, reducing reliance on cross-border supply.

## Conclusions

The medical radioisotope supply chain is a complex and fragile network, and securing its future remains a critical global priority. From literature review and stakeholder interviews, we draw three main conclusions:

Fragility can be divided into a worldwide chronic issue and an acute stressor problem affecting countries without domestic programmes (many LMICs). In North America and Europe acute stressors could mostly be absorbed by the existing (international) network of nuclear reactors, accelerator-based centres, local production, and imports, but this is not the case world-wide. Interviews showed that literature overestimated the risk from acute stressors while highlighting systemic chronic issues including ageing reactors and an underdeveloped workforce.

A geographically distributed production network can reduce both acute and chronic supply fragility and improve access to specific radioisotopes in underserved regions. With a distributed network, a single outage moves from catastrophic loss of access to needed radioisotopes to a redistribution problem. Furthermore, as seen in India and South Africa, LMICs can use this architecture to alleviate import-related risks and cost.

The time to act is now. Novel accelerator concepts are being developed and are in the prototype stage. Regulatory pathways must be discussed and settled in parallel with these demonstrations to avoid licensing holdup. The facility design (targets, hot cells, radiochemistry, etc.) must be in place when the accelerator technology has been fully proven. In parallel, upgrade and replacement strategies for nuclear reactors are happening today.

We call for global coordination and investment across the entire supply chain, encompassing target enrichment, irradiation by accelerator beams or reactor neutrons, separation, transport, radiopharmaceutical preparation, and patient treatment. We argue for complementing and strengthening the existing structures while opening new avenues to increase stability and access for currently underserved regions.

### Panel: Key Takeaways

- Reliance on ageing reactors creates a fragile supply chain potentially exacerbated by short-term stressors.
- The workforce and training pipeline presents concerns to the expansion and sustainability of all levels of nuclear medicine.
- Increased demand for emerging radiopharmaceuticals requires strong production to effectively supply patients in need.
- Increased production complements the established supply chain network increasing reliability and resilience.
- Regional production reduces reliance on international trade, mitigating effects of regulatory differences and international affairs, increasing accessibility and reducing costs, especially in LMICs.

## Declaration of interests

DW, JMC, JM, and DM report institutional research funding or salary support, all paid to MIT, from US Department of Energy awards DE-SC0024914 (which funded this work; see Acknowledgments) and DE-SC0024138, US National Science Foundation awards, a subaward under ARPA-E award DE-AR0002085 via Stony Brook University (DW, JM), and Heising-Simons Foundation grants (DW, JMC). They are developers of the HCHC-XX cyclotron concept described in this Review; MIT holds an MIT-funded contract with Ion Beam Applications SA (IBA) for its concept and feasibility phase, a no-cost agreement for testing at IBA is in preparation, and IBA has made no payments to the authors and holds no rights to the design. CH reports a US NSF Graduate Research Fellowship (DGE-2141064) and MIT-funded research visits to IBA as part of the MIT–IBA studies. PH reports institutional research funding from Siemens Healthineers, Radionetics, and Lilly, and is a co-investigator on Novartis-sponsored clinical trials. PH also reports participation on advisory boards for Telix Pharmaceuticals, Novartis, and BlueEarth Diagnostics, travel support from Telix Pharmaceuticals paid to the author, a signed consulting agreement with ZeEST Bio under which no consultation has yet taken place, institutional receipt of cyclotron facility support from GE Healthcare, and co-founder equity and a board seat in Epitorix, all outside the submitted work. SAE reports institutional research funding from the US National Institutes of Health (K08CA259626), Telix, Sofie, Novartis, and Lilly, and personal consulting fees from Telix, all outside the submitted work. All other authors declare no competing interests.

## Contributors

DW conceived the Series, designed the review, obtained funding, and led the project. DM led the literature search and the stakeholder interviews, curated and analysed the data, and prepared the tables. DM and DW contributed equally to preparing the figures and writing the original draft. JRA and FF drafted sections. PH, SAE, HS, JM, and CH contributed to conception, project coordination, and critical review. JMC contributed to conception, obtained funding, and supervised the work with DW and PH. All authors critically revised the manuscript, had full access to all material, approved the final version, and accept responsibility for the decision to submit.

## Acknowledgments

Contributing interviewees participated in a personal capacity; their participation does not imply an endorsement of the conclusion of this review by them or their institutions. Interviewees were promised confidentiality and their names are omitted without expressed desire to be attributed.

This work was supported by the US Department of Energy, Office of Science, Isotope R&D and Production Program under award number DE-SC0024914 (PI DW, Co-PI JMC). We thank Susan Kayser for proofreading the manuscript for grammar and style. CH was supported by a US National Science Foundation Graduate Research Fellowship under grant number DGE-2141064. Any opinions, findings, and conclusions or recommendations expressed in this material are those of the authors and do not necessarily reflect the views of the US Department of Energy or the National Science Foundation.

The funders had no role in the design of the review, the literature search, the stakeholder interviews, the interpretation of the findings, the writing of the manuscript, or the decision to submit for publication. No author was paid to write this article by any pharmaceutical company or other agency.

## Data sharing

No quantitative datasets were generated for this review. Stakeholder interviews were conducted under the assurance of confidentiality. Interview notes will not be shared. Aggregated, de-identified summaries of interview findings are available from the corresponding author upon reasonable request.

## Declaration of generative AI and AI-assisted technologies in the manuscript preparation process

During the preparation of this work the authors used Claude Opus 5 and Claude Fable 5 (2026) to check spelling and grammar, to comment on the readability and flow of draft sections, to assist with formatting the manuscript to journal requirements, and to support project organisation. Google Translate was used to translate non-English documents. No scientific content, text, figures, tables, or references were generated by AI. After using this tool/service, the authors reviewed and edited the content as needed and take full responsibility for the content of the published article.

# Supplementary Material

## Contents

# Appendix A: Supply Chain Tables

| Year(s) | Reactor or Processing Facility | Duration of Facility Shutdown | Planned or Unplanned Shutdown | Reason for Shutdown | Impact: reduction of $^{99}Mo/^{99m}Tc$ delivery |
|---|---|---|---|---|---|
| 2008 – 2009 | HFR (Netherlands) | 6 months | Unplanned | Gas bubbles detected in main cooling system | 35% |
| 2009 – 2010 | NRU (Canada) | 14 months | Unplanned | Reactor vessel welding repairs | 35% |
| 2010 | HFR | 6 months | Unplanned | Repair of a primary cooling pipework | 35% |
| 2012-2013 | HFR | 8 months | Unplanned | Repair of a primary cooling pipework | 20%* |
| 2013-2014 | Mallinckrodt/**HFR** | 6 months | Unplanned | Issue with control rods | 16%* |
| 2013-2014 | **NTP**/SAFARI-1 (South Africa) | 2 months | Unplanned | Positive pressure in a hot cell caused a leak | 12%* |
| 2015 | HFR | 1 month | Unplanned | Maintenance | 16%* |
| 2015 – 2016 | BR-2 (Belgium) | 16 months | Planned | Major refurbishment | 27%* |
| 2017 – 2018 | **NTP**/SAFARI-1 | 3 months | Unplanned | Regulator shutdown | 12%* |
| 2018 | **NTP**/SAFARI-1 | 6 months | Unplanned | Safety shutdown | 12%* |
| 2018 | BR-2 | 2 weeks | Unplanned | Production issues | 27%* |
| 2020 – 2021 | - | ~10 months** | Unplanned | COVID-19 hindered delivery across borders | 10 – 50%** |
| 2021 | OPAL (Australia) | 2 months | Unplanned | Electronics and mechanical failures | 10%* |
| 2022 | HFR | 2 months | Unplanned | Cooling leak | 22%* |
| 2022 | LVR-15 (Czechia) | 3 weeks | Planned | Maintenance | 11%* |
| 2022 | BR-2 | 2 months | Unplanned | Valve Failure caused delayed restart | 28%* |
| 2022 – 2023 | MARIA (Poland) | 8 months (9 months) | Planned (Unplanned delays) | Major Refurbishment with delayed return to service | 10%* |
| 2024 | OPAL | 6 months | Planned | Major refurbishment | 10%* |
| 2024 | HFR | 3 weeks | Unplanned | Structural pipe deformation | 15%* |
| 2024 | BR-2 | 2 months | Planned | Renovations | 31%* |
| 2024 | MARIA | 1 month | Planned | Routine maintenance | 10%* |
| 2025 | MARIA | 4 months | Unplanned | Licensing halt | 11%* |

*Following the 2009 crisis, the HLG-MR recommendations aided in the mitigation of reactor outage impact

** While the COVID-19 pandemic went on for several years, the global impact on nuclear medicine deliveries varies by country

This table expands on the table provided in the 2016 manuscript Molybdenum-99 for Medical Imaging,[1] and includes the approximate impact of delivery based on the production share of the down reactors.

Notes: HFR = High Flux Reactor, NRU = National Research Universal, NTP = Nuclear Technology Products, SAFARI-1 = South African Fundamental Atomic Research Installation 1, BR-2 = Belgian Reactor 2, OPAL = Open Pool Australian Lightwater reactor.

Data sources: Outages from 2008 – 2016 come from ref [1], Impact was derived from relative outputs from ref [1–7], COVID was taken from ref [8–10]. 2017 and 2018 NTP outages are taken from ref [11], 2018 BR-2 outage from ref [12], 2021 OPAL from ref [13], 2022 HFR from ref [14], 2022 LVR-15 and MARIA outages from ref [15,16], 2022 BR-2 from ref [17], 2024 OPAL from ref [18], 2024 HFR, BR-2 and MARIA outage from ref [19], 2025 MARIA outage from ref [20]

**Table 1. List of $^{99}Mo$ delivery disruptions**

| Project Name | Country | Type | Primary Radioisotopes | Commissioning Year(s) |
|---|---|---|---|---|
| LVR-15 | Czechia | Research Reactor | $^{99}Mo$, $^{153}Sm$, $^{177}Lu$, $^{192}Ir$ | 1957 |
| IEA-R1 | Brazil | Research Reactor | $^{99}Mo$, $^{131}I$, $^{153}Sm$, $^{192}Ir$, $^{198}Au$ | 1957 |
| MNR | Canada | Research Reactor | $^{89}Zr$, $^{90}Y$, $^{125}I$, $^{177}Lu$ | 1959 |
| BRR | Hungary | Research Reactor | $^{32}P$, $^{90}Y$, $^{125}I$, $^{131}I$, $^{153}Sm$, $^{177}Lu$ | 1959 |
| HFR | Netherlands | Research Reactor | $^{90}Y$, $^{99}Mo$, $^{161}Tb$, $^{177}Lu$, $^{192}Ir$ | 1961 |
| BR-2 | Belgium | Research Reactor | $^{99}Mo$, $^{131}I$, $^{177}Lu$, $^{90}Y$ | 1961 |
| JRR-3M | Japan | Research Reactor | $^{60}Co$, $^{192}Ir$, $^{198}Au$ | 1962 |
| Karpov Institute of Physical Chemistry | Russia | Research Reactor | $^{99}Mo$, $^{131}I$, $^{153}Sm$ | 1964 |
| SAFARI-1 | South Africa | Research Reactor | $^{99}Mo$, $^{131}I$, $^{177}Lu$ | 1965 |
| HFIR | United States | Research Reactor | $^{89}Sr$ | 1965 |
| PARR-1 | Pakistan | Research Reactor | $^{32}P$, $^{99}Mo$, $^{131}I$, $^{177}Lu$ | 1965 |
| MURR | United States | Research Reactor | $^{90}Y$, $^{131}I$, $^{177}Lu$, $^{192}Ir$, | 1966 |
| RA-3 | Argentina | Research Reactor | $^{51}Cr$, $^{99}Mo$, $^{131}I$, $^{153}Sm$, $^{192}Ir$ | 1967 |
| TRR | Iran | Research Reactor | $^{32}P$, $^{60}Co$, $^{99}Mo$, $^{131}I$, $^{153}Sm$, $^{166}Ho$, $^{177}Lu$, $^{192}Ir$ | 1967 |
| RIAR | Russia | Research Reactors | $^{89}Sr$, $^{99}Mo$, $^{125}I$, $^{131}I$, $^{153}Sm$, $^{177}Lu$, $^{192}Ir$ | 1961, 1969, 1975, 1983 |
| RECH-1 | Chile | Research Reactor | $^{99m}Tc$, $^{131}I$ | 1974 |
| MARIA | Poland | Research Reactor | $^{32}P$, $^{60}Co$, $^{99}Mo$, $^{131}I$, $^{153}Sm$, $^{166}Ho$, $^{177}Lu$, | 1974 |
| TRR-1/M1 | Thailand | Research Reactor | $^{32}P$, $^{131}I$, $^{153}Sm$ | 1977 |
| WWR SM | Uzbekistan | Research Reactor | $^{32}P$, $^{99m}Tc$, $^{125}I$, $^{131}I$,$^{198}Au$ | 1979 |
| HFETR | China | Research Reactor | $^{60}Co$, $^{99}Mo$, $^{177}Lu$*, $^{192}Ir$, | 1979 |
| TRIGA II | Romania | Research Reactor | $^{131}I$, $^{192}Ir$ | 1980 |
| IRT-1 | Libya | Research Reactor | $^{32}P$, $^{99m}Tc$, $^{131}I$ | 1981 |
| MAYAK | Russia | Research Reactors | $^{32}P$, $^{89}Sr$ | 1979,1988 |
| CANDU (Bruce Power) | Canada | Power Reactors | $^{60}Co$, $^{177}Lu$ | 1984, 1985, 1986, 1987 |
| RSG-GAS | Indonesia | Research Reactor | $^{32}P$, $^{60}Co$, $^{99}Mo$, $^{131}I$, $^{153}Sm$, $^{192}Ir$, $^{198}Au$ | 1987 |
| RP-10 | Peru | Research Reactor | $^{32}P$, $^{99}Mo$, $^{131}I$, $^{153}Sm$, $^{192}Ir$ | 1988 |
| CANDU (Ontario Power Generation) | Canada | Power Reactors | $^{60}Co$, $^{90}Y$*, $^{99}Mo$, $^{177}Lu$* | 1990, 1992 |
| Qinshan Nuclear Power Base | China | Power Reactor | $^{89}Sr$, $^{90}Y$, $^{177}Lu$ | 1994 |
| HANARO | Korea | Research Reactor | $^{32}P$, $^{51}Cr$, $^{60}Co$, $^{99}Mo$, $^{125}I$, $^{131}I$, $^{153}Sm$, $^{165}Dy$, $^{166}Ho$, $^{192}Ir$, $^{198}Au$ | 1995 |
| ETRR-2 | Egypt | Research Reactor | $^{99}Mo$, $^{131}I$ | 1997 |
| BARC | India | Research Reactors | $^{32}P$, $^{60}Co$, $^{64}Cu$, $^{99}Mo$, $^{125}I$, $^{131}I$, $^{192}Ir$, | 1985, 2018 |
| FRM-II | Germany | Research Reactor | $^{99}Mo$*, $^{161}Tb$, $^{166}Ho$, $^{177}Lu$ | 2004 |
| OPAL | Australia | Research Reactor | $^{99}Mo$, $^{125}I$, $^{131}I$, $^{177}Lu$ | 2006 |
| MA-R1 | Morocco | Research Reactor | $^{131}I$ | 2007 |
| CARR | China | Research Reactor | $^{32}P$, $^{60}Co$, $^{99}Mo$,$^{125}I$, $^{131}I$, $^{153}Sm$, $^{192}Ir$, $^{198}Au$ | 2010 |
| CMRR | China | Research Reactor | $^{90}Y$*, $^{99}Mo$,$^{131}I$, $^{161}Tb$, $^{177}Lu$ | 2012 |

**planned expansions*

Data Sources: Research reactors were identified along with their production utilisation in ref [21], non-research reactors, and expansions to production utilisation were identified individually, Russian reactors in ref [22], Canadian reactors in ref [23], the HANARO reactor in ref [24], Indian reactors in ref [25], and Chinese reactors in ref [26–29].

**Table 2. List of commissioned reactors for producing medical radioisotopes**

| Project Name | Country | Type | Primary Radioisotopes | Operational Timeline | Original Timeline |
|---|---|---|---|---|---|
| Jules Horowitz Reactor (JHR) | France | Research Reactor | $^{60}$Co, $^{90}$Y, $^{99}$Mo, $^{103}$Pd, $^{125}$I, $^{153}$Sm, $^{166}$Ho, $^{169}$Er, $^{177}$Lu, $^{186}$Re, $^{192}$Ir | 2032[7] | 2015[4] |
| PALLAS | Netherlands | Research Reactor | $^{90}$Y, $^{99}$Mo, $^{161}$Tb, $^{177}$Lu, $^{192}$Ir | 2032[30] | 2017[4] |
| SHINE Chrysalis | United States | Accelerator | $^{99}$Mo | 2027[7] | 2017[6] |
| Ki-jang Research Reactor (KJRR) | South Korea | Research Reactor | $^{99}$Mo | 2027[31] | 2018[6] |
| Brazilian Multipurpose Reactor (RMB) | Brazil | Research Reactor | $^{99}$Mo | 2031[32] | 2019[6] |
| RA-10 Reactor | Argentina | Research Reactor | $^{99}$Mo, $^{177}$Lu, $^{192}$Ir | 2027[33] | 2019[6] |
| MYRRHA | Belgium | Accelerator-driven system | $^{99}$Mo | 2027[7] | 2022[4] |
| Coquí Pharmaceuticals | United States | Research Reactor | $^{99}$Mo | Stalled[34] | 2022[35] |
| Eden Radioisotopes | United States | Research Reactor | $^{64}$Cu, $^{99}$Mo, $^{161}$Tb, $^{177}$Lu | 2030[36] | 2023[35] |
| NorthStar Medical Radioisotopes | United States | Accelerator | $^{225}$Ac | 2026[37] | 2024[38] |
| Orano ATLab | France | Chemical Extraction | $^{212}$Pb | 2026[39] | 2025[40] |
| Nusano | United States | Accelerator | $^{18}$F, $^{47}$Sc, $^{64}$Cu, $^{67}$Cu, $^{67}$Ga, $^{68}$Ge, $^{82}$Sr, $^{89}$Sr, $^{89}$Zr, $^{99}$Mo, $^{103}$Pd, $^{111}$In, $^{123}$I, $^{124}$I, $^{131}$I, $^{177}$Lu, $^{186}$Re, $^{192}$Ir, $^{211}$At, $^{212}$Pb, $^{225}$Ac, $^{227}$Th | 2026[41] | 2025[42] |
| Research Reactor-2 (RR2) | Nigeria | Research Reactor | $^{99}$Mo | Late 2020s* | 2025[43] |
| Niowave | United States | Accelerators | $^{225}$Ac | 2028[44] | 2028[44] |
| TerraPower Bellwether Facility | United States | Decay Extraction | $^{225}$Ac | 2029[45] | 2029[45] |
| Multi-Purpose Research Reactor (MPR) | South Africa | Research Reactor | $^{99}$Mo | 2033[46] | 2033[46] |
| BARC-IPR | India | Research Reactor | $^{60}$Co, $^{90}$Y, $^{99}$Mo, $^{131}$I, $^{153}$Sm, $^{166}$Ho, $^{177}$Lu, $^{192}$Ir | 2034[47] | 2034[47] |
| NextGEN MURR | United States | Research Reactor | $^{90}$Y, $^{99}$Mo, $^{131}$I, $^{177}$Lu | 2030s[48] | 2030s[48] |

*Originally slated for 2025 operation, RR2 has undergone phase-II review in 2026 without officially adjusting the operational timeline.[49]

**Table 3: Planned and realised expansions of the production sector for 2026 and beyond**

# Appendix B: Interview Structure

Interviews were semi-structured and followed with primarily the same seven questions listed below. Expansions to these questions were individualised to the position and organisation of the interviewee. The interviewees were able to take these questions where they desired to elaborate on their viewpoints freely unless requested to move on for the sake of time. The last question granted the interviewees additional time to complete these thoughts.

1. Are there any isotopes missing from our list that you feel are essential to a comprehensive review?
2. While a few major isotopes ($^{99}$Mo, $^{177}$Lu, $^{131}$I) dominate the supply chain resilience conversation, what other staple isotopes are at risk of a supply chain collapse? How critical would the loss of these lesser-discussed isotopes be for hospital operations?
3. Which isotopes are currently so production-limited that they are stalling clinical trials or creating waitlists?
4. In your view, what has been the most significant failure in the isotope supply chain over the last decade?
5. What points in time have you experienced or witnessed frustration with the access to radioisotopes? Were these frustrations linked to any particular global interruption (COVID, Russia-Ukraine, Brexit, US Tariffs, reactor shutdowns)?
6. Where do you see the weakest link in the supply chain today? Is it material procurement (enriched materials, targets, equipment), production (funding, scale, workforce), cross-border transportation & regulations, or the "last mile" of domestic logistics and radiopharmacy availability?
7. Do you have any final thoughts on supply chain resilience or solutions?

Interviews of clinicians and pharmacists included questions of on-site production and nuclear medicine volume. Interviews of organisations tied to governing bodies covered regional logistics and access to radioisotopes.

# Appendix C: Particle Accelerator Tables

| Class/Type | Energy Range (MeV) | Current Range (mA) | Comments |
|---|---|---|---|
| Electron Accelerators | | | |
| Electron Linac | 30 – 75 | 0·5 – 10 | |
| Rhodotron | 40 | 3·1 | IBA Rhodotron TT300-HE at NorthStar[50] |
| Proton Linacs | | | |
| Dedicated to Isotopes | 25 – 100 | 0·25 – 0·6[51] | LANSCE IPF,[52] KOMAC RI ,[53] Nusano[51] |
| Shared with Physics | 160 – 500 | 0·12 – 0·17 | TRIUMF BL1A[54], Moscow Meson Facility,[22] BNL BLIP[55] |
| Cyclotrons | | | |
| Low Energy Cyclotrons | 9 – 19 | 0.06 – 0.4 | E.g., PET Cyclotrons. Often installed at hospitals. |
| Medium Energy Cyclotrons | 15 – 35 | 0·3 – 1·6 | E.g., SPECT Cyclotrons. Regional commercial class. |
| High Energy Cyclotrons | 35 – 70 | 0·2 – 0·75 | Versatile, often multi-species. Regional commercial class. |

Data sources: An overview is given by refs [56–59]. In addition, electron linac data is taken from refs [60–63], Rhodotron data is taken from refs [50,64], proton linac data is taken from refs[22,51–55], low energy cyclotron data is taken from refs [65,66], medium energy data is taken from refs[65–68], and high energy cyclotron data is taken from refs[65,66,69,70].

**Table 4: Established particle accelerator classes used in the production of radioisotopes.**

| Class/Type | Energy Range (MeV) | Current Range (mA) | Maturity |
|---|---|---|---|
| HCHC-XX/HCDC-XX (High-Current $H_2^+$/$D^+$ Cyclotrons) [71,72] | 3 – 120 $d^+$<br>1·5 – 60 $p^+$ | 5 – 10 | Prototype under construction |
| EcoCyclone (IBA permanent magnet cyclotron) [73] | 18 | 0·25 | Conceptual |
| $H_3^+$ Cyclotrons[74] | 70 – 150 | 0·5 – 1 | Paper study |
| SARAF superconducting $d^+$ linac[75] | 40 | 5 | Under construction, isotopes as future upgrade |
| dCARA ($d^+$ Cyclotron Auto-Resonance Accelerators) [76] | 40 | 25 (peak), 0.0025 (avg) (PoC) | Proof of Concept (PoC) planned |
| ECRIPAC (Electron-Cyclotron Resonance Ion Plasma Accelerator) [77] | 38 – 100 (depending on species) | Not given | Paper study, proof of Concept suggested |
| PW-Laser Driven Acceleration [78] | 10 – 14 | Not given | Demonstrated, but economically not feasible now |
| SYNERGY (Synchrotron-driven photoneutron source)[79] | 10 – 36 (photons) | N/A | Paper study |
| SHINE Chrysalis[80] | 0.3 | 50 – 100 | Facility under construction |
| SORGENTINA-RF[81] | 0.3 | 833 | Design phase |

**Table 5: Particle accelerators under development for potential use in the production of radioisotopes. Note that novel concepts in (subcritical) reactors and accelerator-driven systems are not listed here as they are not particle accelerators.**

| Class/Type | Footprint ($m^2$) | Neutron yield (n/s) | Source |
|---|---|---|---|
| Low Energy Cyclotrons | ~46 | N/A | Company Specifications[65,66,70] + IAEA Facility Guidelines[82] |
| Medium Energy Cyclotrons | ~94 | N/A | Company Specifications[65,66,70] + IAEA Facility Guidelines[82] |
| High Energy Cyclotrons | ~111 | N/A | Company Specifications[65,66,70] + IAEA Facility Guidelines[82] |
| Proton linac (dedicated to isotopes) | ~350-2000 | N/A | Lower: Nusano[51], Upper: KOMAC[53] + shielding and peripherals estimates based on IAEA[82] |
| Electron linacs | ~120 – 175 | N/A | Publication[83] + IAEA Facility Guidelines[82] |
| Rhodotron | ~72 | N/A | Publication[84] + IAEA Facility Guidelines[82] |
| High-current $H_2^+/D^+$ cyclotrons (HCHC-XX/HCDC-XX) (concept) | ~110 – 130 | $9\times10^{12}$ – $8\times10^{15}$ | Publications[72,85] + IAEA Facility Guidelines[82] |
| $D^+$ linac (SARAF-II) § | ~700 | $5\times10^{14}$ – $1\times10^{15}$ | Publications[86] |
| SHINE Chrysalis | ~60 | $1\times10^{14}$ – $1{\cdot}5\times10^{14}$ | Licensing Documents[87] |
| Research reactors | ~350–1050 | $1{\cdot}5\times10^{18}$ – $8\times10^{18}$ | Lower: HIFAR[88], Upper: RJH[89] |
| Sealed D-T generators | ~10-20 | $10^{10}$ – $10^{11}$ | Estimate |

**Table 6: Footprints and neutron yield (where applicable) for the various particle accelerator classes. Only machines that exist or are in the prototype/construction phase are included.**
**Data sources: All cyclotron footprints were estimated based on the manufacturers' specifications. Nusano's dimensions are estimated from the company's press release. KOMAC's dimensions are from published articles. Electron linac dimensions are from published articles. Rhodotron dimensions are from published articles. SARAF-II is based on published material. SHINE is estimated from licensing documents. Research reactors are estimated from published literature. Sealed D-T generators are estimated.**
**The footprint additionally includes the accelerator and peripherals (power supplies, shielding, cryoplants, etc., which are estimates based on facility guidelines given by the IAEA), but not the target stations, radiochemistry, and radiopharmaceutical facilities.**
**§Not currently pursuing medical isotope production but listed as a future application.**